\documentclass{sig-alternate}
\title{Kernel Based Sequential Data Anomaly Detection in Business Process Event Logs}
\numberofauthors{1}
\author{
\alignauthor
Ashish Sureka \\
\affaddr{Software Analytics Research Lab (SARL), India} \\
\email{ashish@iiitd.ac.in}
}

\begin{document}
\maketitle
\begin{abstract}
Business Process Management Systems (BPMS) log events and traces of activities during the execution of a process. Anomalies are defined as deviation or departure from the normal or common order. Anomaly detection in business process logs has several applications such as fraud detection and understanding the causes of process errors. In this paper, we present a novel approach for anomaly detection in business process logs. We model the event logs as a sequential data and apply kernel based anomaly detection techniques to identify outliers and discordant observations. Our technique is unsupervised (does not require a pre-annotated training dataset), employs kNN (k-nearest neighbor) kernel based technique and normalized longest common subsequence (LCS) similarity measure. We conduct experiments on a recent, large and real-world incident management data of an enterprise and demonstrate that our approach is effective.
\end{abstract}
\section{Research Motivation and Aim}
Business Process Management Systems (BPMS), Workflow Management Systems (WMS) and Process Aware Information Systems (PAIS) log events and activities during the execution of a process. Process Mining is a relatively young and emerging discipline consisting of analyzing the event logs from such systems for extracting knowledge such as the discovery of runtime process model (discovery), checking and verification of the design time process model with the runtime process model (conformance analysis) and improving the business process (recommendation and extension) \cite{vanderAalst2004}\cite{vanderAalst2011}. A process consists of cases or traces. A case consists of events. Each event in the event log relates to precisely one case. Events within a case are ordered and have attributes such as activity, timestamp, actor and several additional information such as the cost. The traces and activities in event logs can be modeled as sequential and time-series data. 

Anomaly detection in business process logs is an area that has attracted several researcher's attention. Anomalies are patterns in data that do not conform to a well defined notion of normal behavior. Anomaly detection in business process logs has several applications such as fraud detection, identification of malicious activity and breakdown of the system and understanding the causes of process errors. We conduct a \textit{literature review} of papers closely related to the work presented in this paper. Rogge-Solti et al. propose a Bayesian model that can be automatically inferred from the Petri-Net representation of a business process and is then used to detect non-obvious and temporal anomalies \cite{Solti2014}. Bezerra et al. propose and compares three algorithms for detecting anomalies in logs of process aware systems: threshold, iterative and sampling algorithm. They evaluate the performance of their algorithms on a set of $1500$ artificial logs and demonstrate the effectiveness of their approach \cite{bezerra2008}\cite{Bezerra2013}. 
\begin{figure*}[t]
\centering
\includegraphics[width=0.93\linewidth]{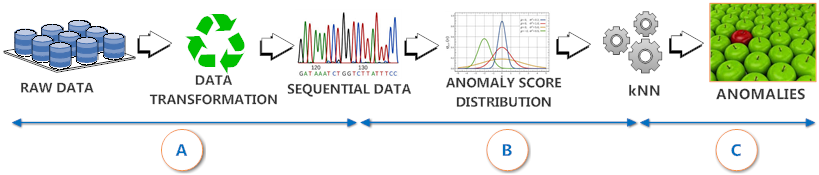}
\caption{Proposed solution approach called as \textit{Nirikshan} consisting of a processing pipeline from raw data transformation to anomaly detection}
\label{fig:framework}
\end{figure*}

The \textit{focus of the study} presented in this paper is on \textsf{anomaly detection in business process logs}. We present a different and fresh perspective to stated problem and our work is motivated by the need to extend the state-of-the-art in the field of techniques anomaly detection in business process event logs. We model the event logs as a sequential data and apply kernel based anomaly detection techniques (which is significant departure from previous approaches) to identify outliers and discordant observations \cite{Chandola2012}. The \textit{research aim and contributions} of the work present in this paper is the following.
\begin{enumerate}
\item To investigate \textsf{kernel based sequential data anomaly detection} based techniques for detecting anomalies and outliers in business process event logs. While there has been work done in the area of anomaly detection in business process logs, to the best of our knowledge, the work presented in this paper is the \textit{first focused study} on application of kernel based sequential data anomaly detection based methods to solve the given problem.
\item To conduct in-depth empirical analysis on real-world dataset and demonstrate the effectiveness of our proposed approach. We conduct experiments on a \textsf{recent, large and real-world incident management data of an enterprise}. The analysis presented in this paper is the \textit{first study} on such a dataset for the application of anomaly detection.    
\end{enumerate}
\section{Experimental Dataset}
\begin{table}[t]
\caption{Actor, Activity and Timestamp for one of the Cases in the Dataset}\label{tab:case} 
\begin{center}
\begin{tabular}{|c|l|c|}
\hline
\scriptsize \textbf{DateStamp} & \scriptsize \centering \textbf{Activity} & \scriptsize \textbf{Group} \\ \hline
\scriptsize 7/1/2013 8:17 & \scriptsize Reassignment & \scriptsize 01 \\ \hline
\scriptsize 4/11/2013 13:41 & \scriptsize Reassignment & \scriptsize 02 \\ \hline
\scriptsize 4/11/2013 13:41 & \scriptsize Update from cust & \scriptsize 02 \\ \hline
\scriptsize 4/11/2013 12:09 & \scriptsize Operator Update & \scriptsize 03 \\ \hline
\scriptsize 4/11/2013 12:09 & \scriptsize Assignment & \scriptsize 03 \\ \hline
\scriptsize 4/11/2013 13:41 & \scriptsize Assignment & \scriptsize 02 \\ \hline
\scriptsize 4/11/2013 13:51 & \scriptsize Closed & \scriptsize 03 \\ \hline
\scriptsize 4/11/2013 13:51 & \scriptsize Caused By CI & \scriptsize 03 \\ \hline
\scriptsize 4/11/2013 12:09 & \scriptsize Reassignment & \scriptsize 03 \\ \hline
\scriptsize 25/09/2013 08:27 & \scriptsize Operator Update & \scriptsize 03 \\ \hline
\end{tabular}
\end{center} 
\end{table}
We conduct our study on large real-world publicly available dataset so that our experiments can be replicated and the results can be used for comparison or benchmarking purposes. The work presented in this paper holds the required \textit{replication standards} ensuring sufficient information for any third party to replicate the results without any additional information from us. We conduct experiments on the publicly available dataset provided by the tenth\footnote{ http://www.win.tue.nl/bpi/2014/challenge} International Workshop on Business Process Intelligence (BPI). Data collection is one of the most important stage in conducting qualitative research and the quality of result obtained depends both on research design and data gathered. The data provided on the BPI workshop website is of high quality as it is peer-reviewed and prepared by experts on the given topic. As an academic, we believe and encourage academic code or software sharing in the interest of improving \textit{openness and research reproducibility}. We release our code and dataset in public domain so that other researchers can validate our scientific claims and use our tool for comparison or benchmarking purposes (and also reusability and extension). Our code and is hosted on GitHub\footnote{currently not mentioned due to blind review policy}[not mentioned due to blind-review policy] which is a popular web-based hosting service for software development projects. 
\begin{figure*}[t]
\centering
\begin{minipage}{0.46\textwidth}
  \centering
  \includegraphics[width=0.93\linewidth]{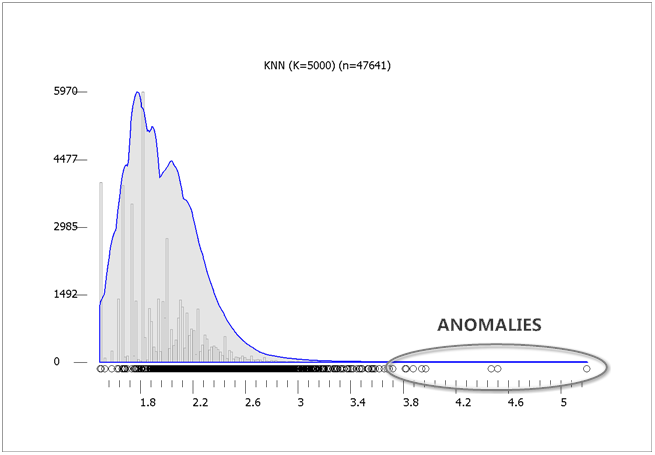}
  \caption{Histogram and kernel density estimate for the anomaly score variable (K value for KNN = $5000$)}
  \label{fig:anomaly5001}
\end{minipage}%
\hspace{1cm}
\begin{minipage}{0.46\textwidth}
\centering
\includegraphics[width=0.93\linewidth]{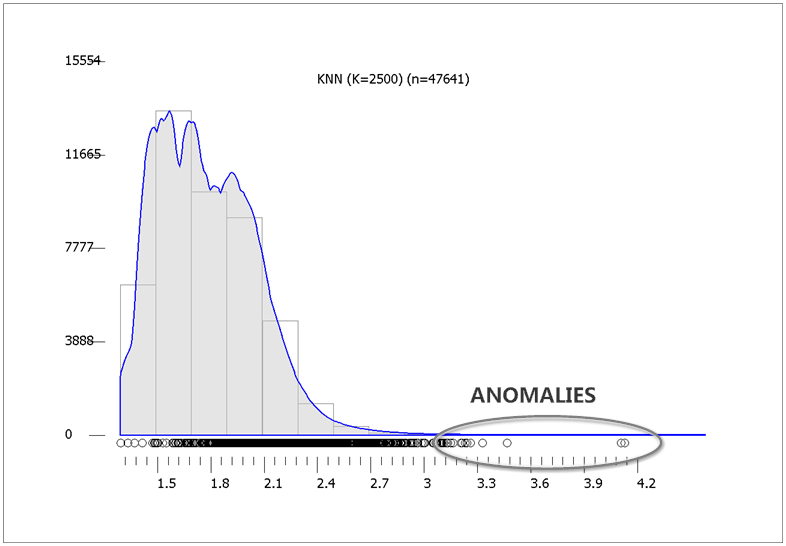}
\caption{Histogram and kernel density estimate for the anomaly score variable (K value for KNN = $2500$)}
  \label{fig:anomaly2501}
\end{minipage}
\end{figure*}
\section{Solution Approach}
Figure \ref{fig:framework} shows the high-level architecture of the proposed solution approach (called as \textit{Nirikshan}) consisting of $3$ steps. The three steps (data transformation, anomaly score distribution computation and application of kNN method) are labeled as $A$, $B$ and $C$ respectively. The Rabobank Group ICT Incident Dataset consists of $46616$ incidents or cases and $466737$ events. The fields in the event-log dataset are: Incident ID, TimeS-stamp, Incident Activity Number, Incident Activity Type, Assignment Group and KM number (a number related to knowledge document). Table \ref{tab:case} shows the Actor, Activity and Timestamp for one of the Cases in the Dataset. The even-log data in Table \ref{tab:case} shows that several activities are performed by various actors during the workflow and process enactment. Table \ref{tab:case} shows that the data has a sequential aspect (is a nature and characteristics of the business process log) and hence we believe techniques for \textsf{anomaly detection for sequences} can be applied to the event log data. While the sequence in the given example is multivariate, in this work, we consider only the activity attribute and model the sequence as univariate. Each case consisting of several events is represented as a sequence of symbols (refer to Phase $A$ of the solution approach in Figure \ref{fig:framework}). Each unique activity is mapped to a symbol. There are $39$ different kinds of activities in the dataset and hence there are $39$ different symbols. Some of the example of activities are: Referred (REF), Problem Closure (PC), OO Response (OOR), Dial-In (DI) and Contact Change (CC). The sequences are of different length. There are a total of $46616$ sequences in the dataset.

\textit{Problem Definition}: In our application, there is no reference or training database available containing only normal sequences. Hence, our task is to detect anomalous sequences (mapped from cases) from an unlabeled database of sequences. The problem is of \textsf{unsupervised anomaly detection}. A formal representation of the problem is \cite{Chandola2012}: Given a set of \textit{n} sequences, $\mathbb{S}$ $=\{S_{1},S_{2},...,S_{n}\}$, find all sequences in $\mathbb{S}$ that are anomalous with respect to rest of $\mathbb{S}$. 

In the unsupervised anomaly detection approach, the entire dataset is treated as a training dataset and a anomaly score is assigned to each sequence with respect to this training dataset (based on the assumption that the training dataset contains few anomalous sequences) \cite{Chandola2012} (refer to Phase $B$ of the solution approach in Figure \ref{fig:framework}). We hypothesize that kernel based techniques (which define an appropriate similarity kernel for the sequences) can be used to detect anomalies for the given dataset and application domain. K-nearest neighbor (kNN) is a well-known and widely used kernel based technique based on a point based anomaly detection algorithm. The main idea behind \textsf{kNN kernel based technique} is to compute the anomaly score for every data point which is equal to the inverse of its similarity to its $k^{th}$ nearest neighbor in the training dataset $\mathbb{S}$ (refer to Phase $C$ of the solution approach in Figure \ref{fig:framework}). Once the anomaly score of each data point is computed, outliers can be detected by identifying the points with high anomaly scores or data points which are $\theta$ (a predefined threshold) standard deviation away from the mean of the anomaly score dataset (assuming the data follows a Gaussian or normal distribution). Kernel based techniques require a similarity kernel. Length of the longest common subsequence (LCS) has been widely used as a similarity kernel for computing the distance (or extent of similarity) between two given sequences. We apply\textsf{ normalized LCS (nLCS) as a similarity measure} between two sequences (which can be of unequal length) $S_{p}$ and $S_{q}$. The formula for nLCS is shown in Equation \ref{eq:LCS}.
\begin{equation}
\label{eq:LCS}
nLCS(S_{p},S_{q})=\frac{ nLCS(S_{p},S_{q})}{\sqrt{|S_{i}||S_{j}|}}
\end{equation}
\section{Empirical Analysis}
We apply kNN based anomaly detection technique with two experimental parameters: k = $5000$ and k = $2500$. We first identify the statistical and density distribution of the anomaly score dataset and check if it has a Gaussian or normal distribution. Figure \ref{fig:anomaly5001} shows the histogram plot dividing the horizontal axis into sub-intervals or bins covering the range of the data from a minimum of $1.49$ to a maximum of $5.19$. The size of the data sample for the histogram and density distribution is the entire population. The solid blue curve is the kernel density estimate which is a generalization over the histogram. We use kernel density estimation to estimate the probability density function of the anomaly score variable. In Figure \ref{fig:anomaly5001}, the data points are represented by small circles on the x-axis. We observe that the data has a Gaussian distribution. The smoothing parameter (bandwidth) for the kernel density estimate in Figure \ref{fig:anomaly5001} is $0.15$. The mean ($\mu$), variance ($\sigma^{2}$) and standard deviation ($\sigma$) for the data is $1.998$, $0.093$ and $0.306$ respectively. We identify anomalies (also called as outliers or discordant observations) by using a standard distance metric to determine how far away each point is from the normal data\footnote{ http://trevorwhitney.com/data\_mining/anomaly\_detection}. The anomalies are marked in the Figure \ref{fig:anomaly5001}. The top $5$ anomaly scores are: $5.196$, $4.516$, $4.467$, $3.968$ and $3.939$. We check how far the data points fall from the mean (also called as the expected value) of the data and how many standard deviations away from the mean that a point is in the dataset. We compute the z-score (calculated using the formula $z=\frac{(x-\mu)}{\sigma})$ for each point which is a measure of how many standard deviations a data point is away from the mean of the data. Any data-point (our interest is on points on the right side of the mean in the given context) that has a z-score higher than $5$ is an outlier, and likely to be an anomaly. The points become more obviously anomalous as the z-score increases above $5$. We found $21$ points with a z-score of more than $5$. The top $5$ z-scores are: $10.451$, $8.230$, $8.070$, $6.439$ and $6.345$. 
\begin{figure}[t]
\centering
\includegraphics[width=0.93\linewidth]{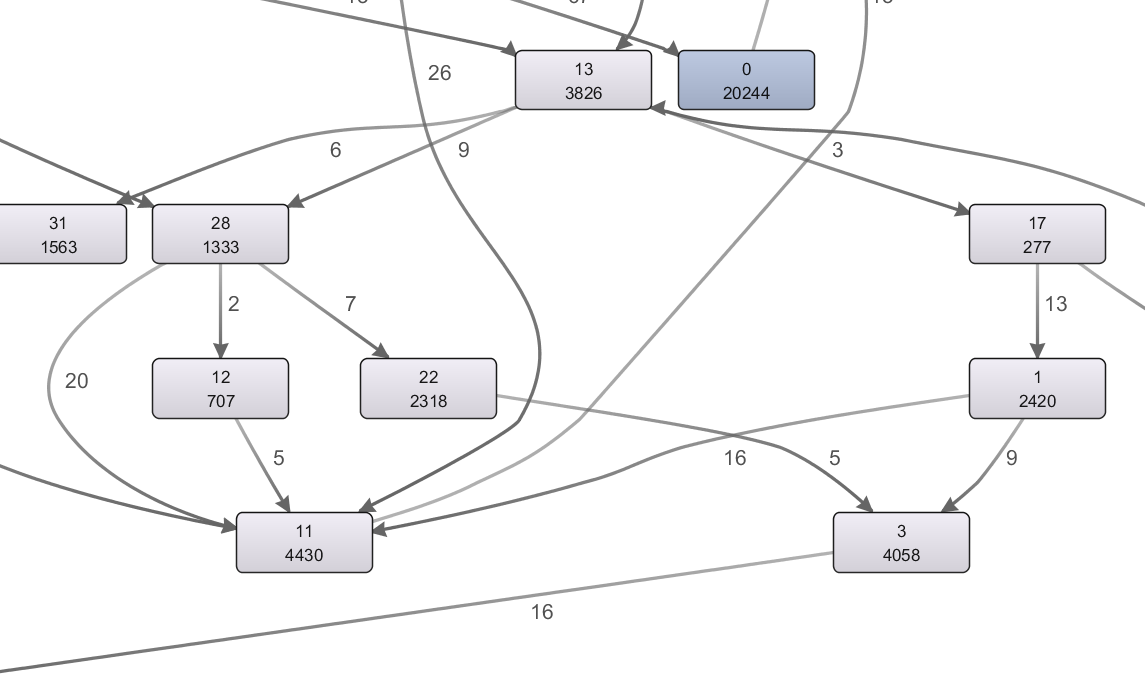}
\caption{Fragment of the discovered process map using DISCO process mining tool at a resolution showing only core transitions}
\label{fig:map} 
\end{figure}
\begin{figure*}[t]
\centering
\begin{minipage}{0.46\textwidth}
  \centering
  \includegraphics[width=0.93\linewidth]{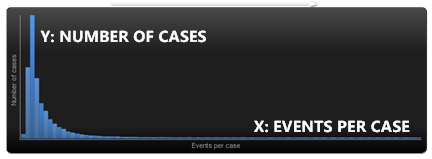}
  \caption{Histogram for events per case}
  \label{fig:epc}
\end{minipage}%
\hspace{1cm}
\begin{minipage}{0.46\textwidth}
\centering
\includegraphics[width=0.93\linewidth]{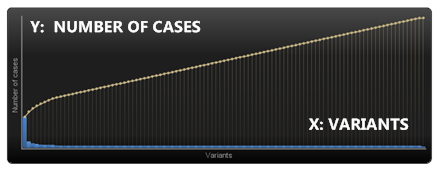}
\caption{Histogram for case variants}
  \label{fig:cv}
\end{minipage}
\end{figure*}

Similarly, we apply the same procedure by setting k = $2500$. Figure \ref{fig:anomaly2501} (k = $2500$) shows the histogram plot and kernel density estimate covering the range of the data from a minimum of $1.29$ to a maximum of $4.58$. We observe that the data has a Gaussian distribution. The mean ($\mu$), variance ($\sigma^{2}$) and standard deviation ($\sigma$) for the data is $1.7692$, $0.0797$ and $0.2823$ respectively. The top $5$ anomaly scores are: $4.582$, $4.127$, $4.106$, $3.464$ and $3.325$. The top $5$ z-scores are: $9.966$, $8.354$, $8.278$, $6.004$ and $5.513$.

We use DISCO\footnote{http://fluxicon.com/disco/} to discover the run-time process model from the given dataset and verify if the anomalous cases identified by our technique matches with the ones extracted by the DISCO tool. Figure \ref{fig:map} shows the fragment of the process model extracted from DISCO (due to limited space it is not possible to display the entire process model). The discovered process model consists of nodes representing the activities in the dataset and directed edges representing the transitions between nodes. The color of the node (and edge thickness) is proportional to the frequency of the activity (darker for more frequency). As shown in Figure \ref{fig:map}, activity label $0$ is dark in color as it has large number of incoming transitions. There are $39$ different activities indexed from $0$ to $38$. The index used is: Caused By CI [$0$], Reopen [$1$], Prob. Work. [$2$], External Vendor Assig. [$3$], Op. Update [$4$], Urgency Change [$5$], Comm. customer [$6$], Impact Change [$7$], Quality Ind. Fixed [$8$], Update [$9$], Anal./Res. [$10$], Desc. Update [$11$], External update [$12$], Pend vendor [$13$], Prob. Closure [$14$], Callback Request [$15$], Update customer [$16$], Notify By Change [$17$], Open [$18$], Dial-in [$19$], Status Change [$20$], Affected CI Change [$21$], Mail Cust [$22$], Referred [$23$], Contact Change [$24$], Reassignment [$25$], Comm. vendor [$26$], Closed [$27$], Resolved [$28$], Quality Indicator Set [$29$], Vendor Ref Change [$30$], Quality Indicator [$31$], Vendor Ref [$32$], Incident repr [$33$], External Vendor Reass [$34$], Assig [$35$], OO Response [$36$], alert stage 1 [$37$], Service Change [$38$].
\begin{table*}[t]
\caption{Anomalies extracted from the proposed solution approach (kNN kernel based techniqie)}\label{tab:anom} 
\begin{center}
\begin{tabular}{|c|p{16cm}|}
\hline
\scriptsize 1 & \scriptsize 35;35;35;35;9;25;25;25;1;27;1;4;4;9;4;9;9;9;9;9;25; 25;25;11;25;25;25;4;20;9;4;35;35;35;35;35;35;25;35;35;35; 27;\\ \hline
\scriptsize 2 & \scriptsize 27; 6; 27; 9; 4; 4; 16; 6; 27; 16; 27; 27; 16; 6; 27; 27; 27; 6; 4; 27; 27; 4; 4; 17; 27; 27; 27; 27; 4; 4; 4; 16; 27; 6; 27; 16; 18; 6; 25; 9; 35; 20; 35; 6; 20; 27; 27; 27; 27; 27; 27; 27; 27; 27; 4; 11; 27; 27; 27; 27; 27; 24; 6; 16; 4; 16; 4; 4; 27; 27; 27; 16; 27; 27; 27; 27; 16; 16; 2; 16; 6; 16; 27; 35; 25; 27; 27; 4; 35; 4; 6; 35; 27; 27; 4; 6; 27; 27; 27; 35; 6; 27; 27; 20; 35; 16; 35; 4; 4; 16; 27; 27; 16; 27; 27; 16; 27; 27; 4; 16; 4; 27; 5; 4; 27; 35; 6; 4; 25; 16; 16; 16; 27; 35; 4; 27; 16; 27; 27; 4; 16; 16; 27; 6; 6; 27; 16; 27; 6; 27; 27; 27; 27; 4; 9; 27; 16; 27; 6; 6; 6; 9; 4; 17; 27; 27; 27; 27; 6; 27; \\ \hline
\scriptsize 3 & \scriptsize 20; 27; 0; 35; 32; 25; 18; 3; 20 \\ \hline
\scriptsize 4 & \scriptsize 25; 25; 11; 18; 27; 0; \\ \hline
%\scriptsize 5 & \scriptsize 25; 16; 18; 35; 25; 25; 4; 4; 35; 16; 4; 35; 35; 25; 9; 4; 35; 4; 4; 35; 4; 4; 4; 4; 35; 35; 35; 4; 35; 35; 35; 25; 35; 4; 35; 25; 35; 35; 25; 35; 25; 4; 25; 35; 25; 4; 35; 25; 16; 4; 4; 35; 35; 35; 25; 0; 4; 35; 16; 27; 25; 35; 9; 25; 25; 4; 35; 25; 35; 4; 25; 35; 4; 25; 35; 35; 25; 35; 25; 16; 35; 35; 16; 35; 4; 25; 4; 35; 35; 35; 35; 25; 25; 4; 4; 4; 4; 25; 25; 35; 4; 9; 4; 4; 35; 25; 35; 4; 35; 25; 25; 16; 4; 25; 35; 4; 35; 35; 35; 4; 25; 25; 4; 35; 20; 25; 35; 25; 35; 4; 25; 4; 35; 25; 4; 4; 35; 25; 25; 35; 35; 35; 35; 9; 35; 4; 35; 35; 35; \\ \hline
\scriptsize 5 & \scriptsize 0; 27; \\ \hline
%\scriptsize 7 & \scriptsize 0; 8; 27; \\ \hline
\hline
\end{tabular}
\end{center} 
\end{table*}
Figure \ref{fig:epc} shows the distribution of the number of events per case. Figure \ref{fig:cv} shows the distribution of the case invariants. Both the distributions displayed in Figure \ref{fig:epc} and \ref{fig:cv} are skewed as one of the tails is longer than the other. Both the distributions has a positive skew as the long tail is in the positive direction. Distribution in Figure \ref{fig:epc} reveals that most of the cases has few events whereas a small number of cases consists of large number of events. Similarly, the distribution in Figure \ref{fig:cv} indicates that the mean and median of the case variants is greater than the mode and the dataset consists of a long of small number of case invariants. Table \ref{tab:anom} shows $5$ of the top $10$ anomalies extracted by our approach. We validate it with the output of the DISCO infrequent case variants.
\section{Conclusions}
We present a technique to detect anomalies from business process event logs. We apply KNN kernel based sequential anomaly detection based method and conduct experiments on real-world dataset. We validate the effectiveness of the proposed approach and conclude that kernel based sequential data anomaly detection techniques can be effectively applied for the domain of extracting outliers from business process event logs. We learn that similarity kernel used (for example, nLCS) in the proposed technique and the value of the kNN parameter ($K$) has an effect on the outcome. 
\bibliographystyle{abbrv}
\bibliography{comad2014}
\end{document}